\documentstyle [12pt]{article}
\def\e{{\rm e}}

\def\tr{{\rm tr}}

\def\coth{\mathop{\rm coth}\nolimits}

\def\one#1{#1^{\raise5pt\hbox{$\scriptstyle\!\!\!\!1$}}\,{}}
\def\two#1{#1^{\raise5pt\hbox{$\scriptstyle\!\!\!\!2$}}\,{}}
\def\three#1{#1^{\raise5pt\hbox{$\scriptstyle\!\!\!\!3$}}\,{}}

\def\phi{\varphi}

\def\a{\alpha}
\def\b{\beta}
\def\d{\delta}

\def\s{\sigma}
\def\beq{\begin{equation}}
\def\eeq{\end{equation}}
\def\be{\begin{displaymath}}
\def\ee{\end{displaymath}}
\def\bm{\left(\begin{array}}
\def\em{\end{array}\right)}
\def\bds{\begin{description}}
\def\eds{\end{description}}
\def\?{(?)\marginpar{|?}}

\def\half{\frac{1}{2}}

\newtheorem{pr}{Proposition}

\def\g{\gamma}

\def\H{{\cal H}}
\def\suma{\sum_{\a=1}^N}
\def\sumab{\sum_{\a\neq\b}}
\def\id{\hbox{{1}\kern-.25em\hbox{l}}}
\def\carre#1#2#3#4{[\!\phantom{|}^{#1#2}_{#3#4}]}
\def\Carre#1#2#3#4#5#6{[\!\phantom{|}^{#1#2#3}_{#4#5#6}]}
\begin{document}
\title{Dynamical $r$-matrices for the
       Elliptic Calogero-Moser Model}
\author{E. K. Sklyanin \thanks{On leave from Steklov Mathematical Institute,
Fontanka 27, St.Petersburg 191011, Russia.}\\
\normalsize
Laboratoire de Physique Th\'eorique et Hautes Energies
\thanks{Laboratoire associ\'e au CNRS $n^o$ 280}\\
\normalsize
Universit\'e Pierre et Marie Curie,
Tour 16, $1^{er}$ \'etage, 4 place Jussieu,\\
\normalsize
75252 Paris cedex 05, France
        }
\date{August 12, 1993}
\maketitle
\vskip-9.5cm
\hskip12.0cm
\sf LPTHE-93-42\rm
\vskip.2cm
\hskip11.4cm
\sf hep-th/9308060 \rm
\vskip8.8cm

{\bf Abstract.} For the integrable $N$-particle Calogero-Moser system with
elliptic potential it is shown that the Lax operator found by Krichever
possesses a classical $r$-matrix structure. The $r$-matrix is a natural
generalisation of the matrix found recently by Avan and Talon
(hep-th/9210128) for the trigonometric potential. The $r$-matrix depends
on the spectral parameter and only half of the dynamical variables
(particles' coordinates).
It satisfies a generalized Yang-Baxter equation involving another dynamical
matrix.

%\newpage
\section{Introduction}

The elliptic Calogero-Moser model is the system of $N$ one-dimensional
particles interacting via two-particle potential
\beq
V(q_{\a\b})\equiv\wp(q_{\a\b}), \qquad q_{\a\b}\equiv  q_\a-q_\b,
\label{eq:defV}
\eeq
$\wp$ being the Weierstrass function \cite{Bateman} with the periods
$2\omega_1$, $2\omega_2$.
In terms of the canonical variables $p_\a$, $q_\a$ ($\a=1,\ldots,N$)
\beq
 \{p_\a, p_\b\}=\{q_\a, q_\b\}=0, \qquad \{p_\a, q_\b\}=\d_{\a\b}
\eeq
the Hamiltonian of the system is expressed as
\beq
  \H=\suma p^2_\a + \sumab V(q_{\a\b}),
\label{eq:defH}
\eeq
(we deliberately omit a coupling constant which always can be removed
by rescaling).

The Hamiltonian (\ref{eq:defH}) with the potential (\ref{eq:defV}) is known
to be completely integrable \cite{Cal75,Cal76,Moser,OlP81}. The most effective
way
to construct
the complete set of commuting integrals of motion is to represent them
as the spectral invariants of a Lax operator --- certain $N\times N$
matrix depending on the dynamical variables and, may be,  on an additional
parameter $u$ called spectral.  As shown in \cite{BV90}, for the commutativity
of the spectral invariants $\tr L(u)^n$ of the Lax operator it is necessary
and sufficient that the Poisson bracket
$\{L^{\a_1}_{\b_1}(u), L^{\a_2}_{\b_2}(v)\}$
could be represented in the commutator form
\begin{eqnarray}
 \{L^{\a_1}_{\b_1}(u), L^{\a_2}_{\b_2}(v)\}=\sum_{\g_1\g_2}&&
 r^{\a_1\a_2}_{\g_1\b_2}(u,v)L^{\g_1}_{\b_1}(u)
-L^{\a_1}_{\g_1}(u)r^{\g_1\a_2}_{\b_1\b_2}(u,v) \nonumber \\
&&-r^{\a_2\a_1}_{\g_2\b_1}(v,u)L^{\g_2}_{\b_2}(v)
+L^{\a_2}_{\g_2}(v)r^{\g_2\a_1}_{\b_2\b_1}(v,u)
\label{eq:LLind}
\end{eqnarray}
or, using the notation \cite{FT}
$$
 L^{(1)}\equiv L\otimes \id, \qquad L^{(2)}\equiv \id\otimes L,
$$
as
\beq
 \{L^{(1)}(u), L^{(2)}(v)\}=[r^{(12)}, L^{(1)}(u)]-[r^{(21)}, L^{(2)}(v)]
\label{eq:LL}
\eeq
where $r^{(12)}$ is an $N^2\times N^2$ matrix depending, generally
speaking, on the dynamical variables, and $r^{(21)}$ is
\beq
  r^{(21)}(u,v)\equiv P r^{(12)}(v,u) P,
\label{eq:defr21}
\eeq
$P$ being the permutation: $Px\otimes y=y\otimes x$.

In contrast with the well-studied case of purely numerical $r$-matrices
\cite{FT}, no general theory of dynamical $r$-matrices exists at the  moment,
apart few concrete examples and observations
\cite{BV90,Maillet,FA,AT,ABT,BB,EEKT}.
Still, the collection of examples is rather sparse, and any new example
of dynamical $r$-matrix could possibly contribute to better understanding
of their algebraic and geometric nature.

In \cite{AT} the dynamical $r$-matrix has been calculated for
 an $L$-operator without spectral parameter corresponding to
Calogero-Moser model with the trigonometric potential $V(q)=1/\sin^2q$
or $1/\sinh^2q$. In \cite{ABT} an elegant interpretation of the $r$-matrix
was found in terms of hamiltonian reduction.  However, the $L$-operator
used in \cite{AT,ABT} lacks the spectral parameter. Though the absence
of the spectral parameter is not important for the construction of
commuting integrals of motion, the spectral parameter is indispensable
for integrating  equations of motion and constructing
the action-angle variables. The proper way of introducing a
spectral parameter into $L$-operator was found by Krichever \cite{Krich}.

In the present paper the dynamical $r$-matrix is found for the
Krichever's $L$-operator corresponding to the elliptical potential
(\ref{eq:defV}). The $r$-matrix found previously in \cite{AT} is shown
to be a degenerate case of our $r$-matrix. We derive also an equation
satisfied by the dynamical $r$-matrix which generalises the classical
Yang-Baxter equation satisfied by numerical $r$-matrices. Our equation
has a slightly more general form that the equation for a
dynamical $r$-matrix proposed recently in \cite{EEKT}.

\section{$L$-operator and dynamical $r$-matrix}

Krichever's $L$-operator \cite{Krich} is the $N\times N$ matrix
given by the formula
\beq
 L^\a_\b(u)=p_\a \d_{\a\b}+iQ(u,q_{\a\b})(1-\d_{\a\b}),
\label{eq:defLind}
\eeq
or, in terms of the basic matrices $E^\a_\b$
$$
(E^\a_\b)^{\a^\prime}_{\b^\prime}=\d_{\a\a^\prime}\d_{\b\b^\prime},
$$
by
\beq
 L(u)=\suma  p_\a E^\a_\a+i\sumab Q(u,q_{\a\b})E^\a_\b,
\label{eq:defL}
\eeq
where $Q(u,q)$ is expressed in terms of Weierstrass $\s$ functions
\beq Q(u,q)\equiv\frac{\s(u+q)}{\s(u)\s(q)}
\label{eq:defQ}
\eeq

For what follows, it suffices to know that $\s(z)$ is the entire function
defined by the product \cite{Bateman}
$$
 \s(z)=z\prod_{m,n\neq0}\left(1-\frac{z}{\omega_{mn}}\right)
     \exp\left[\frac{z}{\omega_{mn}}+
         \half\left(\frac{z}{\omega_{mn}}\right)^2\right]
$$
where $\omega_{mn}=2m\omega_1+2n\omega_2$, $2\omega_{1,2}$ being a pair of
periods.
The $\zeta$ and $\wp$ functions are introduced through the derivatives of $\s$
\beq
 \zeta(z)=\frac{\s^\prime(z)}{\s(z)}, \qquad \wp(z)=-\zeta^\prime(z)
\eeq

The translations of $z$ by $2\omega_{1,2}$ and the reflection $z\rightarrow-z$
act on the functions as follows
$$
 \begin{array}{c}
 \s(z+2\omega_l)=-\s(z)\exp[2(z+\omega_l)\zeta(\omega_l)], \cr  \cr
 \zeta(z+2\omega_l)=\zeta(z)+2\zeta(\omega_l), \qquad
 \wp(z+2\omega_l)=\wp(z).
 \end{array}
$$

\beq
 \s(-z)=-\s(z), \qquad \zeta(-z)=-\zeta(z), \qquad \wp(-z)=\wp(z).
\label{eq:asymszp}
\eeq

In the vicinity of $z=0$ the Weierstrass functions have the expansions
$$
 \s(z)=z+O(z^5), \qquad \zeta(z)=z^{-1}+O(z^3), \qquad
               \wp(z)=z^{-2}+O(z^2).
$$

The spectral invariants $\tr L(u)^n$, $n=1,\ldots,N$ are independant
commuting quantities \cite{OlP81}.
For instance, the total momentum ${\cal P}=\suma p_\a$ and the Hamiltonian
$\H$ (\ref{eq:defH}) are obtained from
\beq
 \tr L(u)={\cal P}, \qquad \tr L(u)^2=\H-V(u),
\eeq
the last equality following from the identity
\beq
 Q(u,q_{\a\b})Q(u,q_{\b\a})=V(u)-V(q_{\a\b})
\eeq
which follows, in turn, from
\beq
 -\frac{\s(u-v)\s(u+v)}{\s^2(u)\s^2(v)}=\wp(u)-\wp(v),
\eeq
see \cite{Bateman}.

\begin{pr}
 For the $L$-operator (\ref{eq:defL}) the identity (\ref{eq:LL}) holds
with the following matrices $r^{(12)}$ et $r^{(21)}$
\beq
        r^{(12)}(u,v)=a\suma  E_{\a\a}^{\a\a}
         +\sumab c_{\a\b}E^{\a\b}_{\b\a}
         +\sumab d_{\a\b}(E^{\a\a}_{\a\b}+E^{\b\a}_{\b\b})
\label{eq:defr}
\eeq
where
$$
 E^{\a_1\a_2}_{\b_1\b_2}\equiv E^{\a_1}_{\b_1}\otimes E^{\a_2}_{\b_2}
$$

\beq
 \begin{array}{c}
  a=r^{\a\a}_{\a\a}=-\zeta(u-v)-\zeta(v), \qquad
  c_{\a\b}=r^{\a\b}_{\b\a}=-Q(u-v,q_{\a\b}), \qquad \cr  \cr
  d_{\a\b}=r^{\a\a}_{\a\b}=r^{\b\a}_{\b\b}=
      -\half Q(v,q_{\a\b}),
 \end{array}
\label{eq:defacd}
\eeq
$r^{(21)}$ being defined by (\ref{eq:defr21})
\end{pr}

{\bf Proof.} Notice, first  of all, that, due to (\ref{eq:defacd}), the
tensor $r^{\a_1\a_2}_{\b_1\b_2}$ has nonzero components only for combinations
of four indices $\carre{\a_1}{\a_2}{\b_1}{\b_2}$ having no more than
{\it two} different indices. This property of $r$ simplifies greatly
verification
of the identity (\ref{eq:LL}) since, in this case, it suffices to consider
only the combinations $\carre{\a_1}{\a_2}{\b_1}{\b_2}$ in (\ref{eq:LLind})
having no more than {\it three} different indices. The obvious symmetry
of (\ref{eq:defacd}) with respect to the permutations of indices allows
to reduce the task to the verification of (\ref{eq:LLind}) for only
14 combinations: $\carre1111$, $\carre1112$, $\carre1121$, $\carre1211$,
$\carre1222$, $\carre1212$, $\carre1221$, $\carre1122$, $\carre1213$,
$\carre1232$, $\carre1123$, $\carre1233$, $\carre1231$, $\carre1223$.

Substituting (\ref{eq:defr}) into (\ref{eq:LLind})
results for the above combinations  of indices in the equalities
\begin{eqnarray}
 \carre1111:\quad\{L^1_1(u),L^1_1(v)\}&=&0, \nonumber \\
 \carre1112:\quad\{L^1_1(u),L^1_2(v)\}&=&
     -L^1_2(u)c_{21}(u-v)-a(v,u)L^1_2(v),  \nonumber \\
 \carre1121:\quad\{L^1_2(u),L^1_1(v)\}&=&
     a(u,v)L^1_2(u)+L^1_2(v)c_{21}(v-u), \nonumber \\
 \carre1211:\quad\{L^1_1(u),L^2_1(v)\}&=&
     c_{12}(u-v)L^2_1(u)+L^2_1(v)a(v,u), \nonumber \\
 \carre1222:\quad\{L^1_2(u),L^2_2(v)\}&=&
     -L^1_2(u)a(u,v)-c_{21}(v-u)L^1_2(v), \nonumber \\
 \carre1212:\quad\{L^1_1(u),L^2_2(v)\}&=&0, \nonumber \\
 \carre1221:\quad\{L^1_2(u),L^2_1(v)\}&=&
      c_{12}(u-v)(L^2_2(u)-L^1_1(u))
      +c_{21}(v-u)(L^2_2(v)-L^1_1(v)), \nonumber \\
 \carre1122:\quad\{L^1_2(u),L^1_2(v)\}&=&0, \nonumber \\
 \carre1213:\quad\{L^1_1(u),L^2_3(v)\}&=&0, \nonumber \\
 \carre1232:\quad\{L^1_3(u),L^2_2(v)\}&=&0, \nonumber \\
 \carre1123:\quad\{L^1_2(u),L^1_3(v)\}&=&
      d_{13}(v)L^1_2(u)-d_{12}(u)L^1_3(v), \nonumber \\
 \carre1233:\quad\{L^1_3(u),L^2_3(v)\}&=&
      -L^1_3(u)d_{23}(v)+L^2_3(v)d_{13}(u), \nonumber \\
 \carre1231:\quad\{L^1_3(u),L^2_1(v)\}&=&
      d_{21}(v)L^1_3(u)+c_{12}(u-v)L^2_3(u) \nonumber \\
      &&+L^2_1(v)d_{13}(u)+L^2_3(v)c_{31}(v-u), \nonumber \\
 \carre1223:\quad\{L^1_2(u),L^2_3(v)\}&=&
      -L^1_2(u)d_{23}(v)-L^1_3(u)c_{32}(u-v) \nonumber \\
      &&-d_{12}(u)L^2_3(v)-c_{21}(v-u)L^1_3(v). \nonumber
\end{eqnarray}

Using then the substitution (\ref{eq:defLind}) one obtains the trivial
identities $0=0$ for
$\carre{\a_1}{\a_2}{\b_1}{\b_2}=\carre1111$, $\carre1212$, $\carre1122$,
$\carre1213$, and $\carre1232$.
After the substitutions (\ref{eq:defacd}) the same trivial identity
is obtained for
$\carre{\a_1}{\a_2}{\b_1}{\b_2}=\carre1123$, and $\carre1233$.

In the same manner, the case $\carre1221$ is reduced to the equality
$$
 0=(p_1-p_2)[Q(u-v,q_{12})+Q(v-u,q_{21})]
$$
which follows from the identity
\beq
  Q(-u,-q)=-Q(u,q)
\label{eq:asymQ}
\eeq
resulting from (\ref{eq:defQ}) and (\ref{eq:asymszp}).

Analogously, for $\carre{\a_1}{\a_2}{\b_1}{\b_2}=\carre1231$ and $\carre1223$
one obtains, respectively, the identities
\begin{eqnarray}
 0&=&Q(v,q_{21})Q(u,q_{13})+Q(u-v,q_{12})Q(u,q_{23})+Q(v,q_{23})Q(v-u,q_{31})
              \nonumber \\
 0&=&Q(u,q_{12})Q(v,q_{23})+Q(u,q_{13})Q(u-v,q_{32})+Q(v-u,q_{21})Q(v,q_{13})
              \nonumber
\end{eqnarray}
which, after substitution of (\ref{eq:defQ}), are reduced to the standard
three-term quartic identity for $\s$-function  \cite{Halphen}
\begin{eqnarray}
 &&  \s(x-y)\s(x+y)\s(z-t)\s(z+t) \nonumber \\
 &+& \s(y-z)\s(y+z)\s(x-t)\s(x+t) \nonumber \\
 &+& \s(z-x)\s(z+x)\s(y-t)\s(y+t)=0.
\label{eq:3term}
\end{eqnarray}

Similarly, the remaining cases $\carre{\a_1}{\a_2}{\b_1}{\b_2}=\carre1112$,
$\carre1121$, $\carre1211$, and $\carre1222$ lead, respectively, to the
identities
\begin{eqnarray}
 {[}\zeta(v+q_{12})-\zeta(q_{12}){]}Q(v,q_{12})&=&Q(u-v,q_{21})Q(u,q_{12})
   +{[}\zeta(v-u)+\zeta(u){]}Q(v,q_{12}), \nonumber \\
 {[}\zeta(u+q_{12})-\zeta(q_{12}){]}Q(u,q_{12})&=&Q(v-u,q_{21})Q(v,q_{12})
   +{[}\zeta(u-v)+\zeta(v){]}Q(u,q_{12}), \nonumber \\
 {[}\zeta(v+q_{21})-\zeta(q_{21}){]}Q(v,q_{21})&=&Q(u-v,q_{12})Q(u,q_{21})
   +{[}\zeta(v-u)+\zeta(u){]}Q(v,q_{21}), \nonumber \\
 {[}\zeta(u+q_{12})-\zeta(q_{12}){]}Q(u,q_{12})&=&Q(v-u,q_{21})Q(v,q_{12})
   +{[}\zeta(u-v)+\zeta(v){]}Q(v,q_{12}), \nonumber
\end{eqnarray}
which are reduced to another standard identity \cite{Halphen}
\beq
 \zeta(x+z)-\zeta(x-z)+\zeta(y-z)-\zeta(y+z)=
\frac{\s(2z)\s(x+y)\s(x-y)}{\s(x+z)\s(x-z)\s(y+z)\s(y-z)}
\label{eq:4term}
\eeq
being a derivative form of (\ref{eq:3term}).

\section{Degenerate cases}

The degenerate cases discussed below correspond to the  infinite values of
one or both of the periods $\omega_{1,2}$. The corresponding potential
$V(q)$ becomes respectively a trigonometric/hyperbolic or rational function
of $q$.
According to \cite{OlP81} we shall mark these cases as I ($V(q)=q^{-2}$),
II ($V(q)=\sinh^{-2}q$), and III ($V(q)=\sin^{-2}q$), the case IV
corresponding to the original elliptic potential $V(q)=\wp(q).$

In the case II the periods are $\omega_1=\infty$, $\omega_2=\pi i/2$,
the Weierstrass functions degenerate into \cite{Bateman}
\beq
 \s(z)=\sinh z\exp\left[-\frac{z^2}{6}\right], \qquad
 \zeta(z)=\coth z-\frac{z}{3}, \qquad
 \wp(z)=\frac{1}{\sinh^2z}+\frac{1}{3}
\eeq
and $Q(u,q)$, see (\ref{eq:defQ}), into
\beq
 Q(u,q)=\frac{\sinh(u+q)}{\sinh u\sinh q}\exp\left[-\frac{uq}{3}\right]
\eeq

One can simplify the corresponding $L$ operator (\ref{eq:defL}) removing the
factors $\e^{-uq_{\a\b}/3}$ by means of the similarity transform
\beq
 L(u)\longrightarrow L^\prime(u)=W(u)L(u)W^{-1}(u), \qquad
      W^\a_\b(u)=\d_{\a\b}\e^{uq_\a/3}.
\eeq

Calculating the Poisson bracket $\{L^{\prime(1)}(u),L^{\prime(2)}(v)\}$
\begin{eqnarray}
 \{L^{\prime(1)},L^{\prime(2)}\}&=&
   \{W^{(1)}L^{(1)}(W^{(1)})^{-1},W^{(2)}L^{(2)}(W^{(2)})^{-1}\}
  \nonumber \\
&=&W^{(1)}W^{(2)}\{L^{(1)},L^{(2)}\}(W^{(1)})^{-1}(W^{(2)})^{-1} \nonumber \\
&&+W^{(2)}\{W^{(1)},L^{(2)}\}(W^{(2)})^{-1}L^{(1)}(W^{(1)})^{-1} \nonumber \\
 && -W^{(2)}W^{(1)}L^{(1)}(W^{(1)})^{-1}\{W^{(1)},L^{(2)}\}
(W^{(1)})^{-1}(W^{(2)})^{-1} \nonumber \\
&&+W^{(1)}\{L^{(1)},W^{(2)}\}(W^{(1)})^{-1}L^{(2)}(W^{(2)})^{-1} \nonumber \\
&&  -W^{(1)}W^{(2)}L^{(2)}(W^{(2)})^{-1}\{L^{(1)},W^{(2)}\}
(W^{(2)})^{-1}(W^{(1)})^{-1}
\label{eq:LLmod}
\end{eqnarray}
(we have omitted the arguments $u$ and $v$ which are easy to  restore)
and using the identities
\begin{eqnarray}
 W^{(2)}\{W^{(1)},L^{(2)}\}(W^{(2)})^{-1}(W^{(1)})^{-1}&=&-\frac{u}{3}E
                                  \nonumber \\
 W^{(1)}\{L^{(1)},W^{(2)}\}(W^{(1)})^{-1}(W^{(2)})^{-1}&=&\frac{v}{3}E
                                  \nonumber
\end{eqnarray}
where
$$
 E=\suma E^{\a\a}_{\a\a}
$$
one can show that $L^\prime$ satisfies again the relation (\ref{eq:LL})
with the modified $r$ matrix which is also described by the formulas
(\ref{eq:defr}) and (\ref{eq:defacd}) where one should replace
the Weierstrass functions with
\beq
 \s(z)\longrightarrow\sinh z, \qquad
 \zeta(z)\longrightarrow\coth z, \qquad
 \wp\longrightarrow \frac{1}{\sinh^2z}
\label{eq:szp-II}
\eeq
and, respectively,
\beq
    Q(u,q)\longrightarrow \frac{\sinh(u+q)}{\sinh u\sinh q}=
             \coth u+\coth q.
\label{eq:Q-II}
\eeq

Quite analogously, the case III, corresponding to the periods
$\omega_1=\pi/2$ and $\omega_2=i\infty$, is reduced to the substitutions
$$
  \s(z)\longrightarrow\sin z, \qquad
 \zeta(z)\longrightarrow\cot z, \qquad
 \wp\longrightarrow \frac{1}{\sin^2z},
$$
and
$$
    Q(u,q)\longrightarrow \frac{\sin(u+q)}{\sin u\sin q}=
           \cot u+\cot q.
$$

The case I corresponds to the periods $\omega_1=\infty$ and $\omega_2=i\infty$
and to the substitutions
\beq
  \s(z)\longrightarrow z, \qquad
 \zeta(z)\longrightarrow\frac{1}{z}, \qquad
 \wp\longrightarrow \frac{1}{z^2},
\label{eq:szp-I}
\eeq
and
\beq
    Q(u,q)\longrightarrow \frac{u+q}{uq}=
           \frac{1}{u}+\frac{1}{q}.
\label{eq:Q-I}
\eeq

In \cite{AT} the $r$ matrices were calculated for the I, II and III type
potentials and the $L$ operators containing no spectral parameters.
For the case I the $L$ operator used in \cite{AT} has the form
(\ref{eq:defL}) with $Q(u,q)$ replaced by $q^{-1}$. The corresponding $r$
matrix has the structure (\ref{eq:defr}) with
\beq
 a=0, \qquad c_{\a\b}=-\frac{1}{q_{\a\b}}, \qquad
      d_{\a\b}=-\frac{1}{2q_{\a\b}}
\eeq

It is easy to see that the above result can be reproduced from ours
(\ref{eq:szp-I}), (\ref{eq:Q-I}) by taking the limits
$u,v\longrightarrow\infty$. The cases II and III need, however, some tricks.

It is sufficient to consider the case II since the case III is obtained by the
obvious change of the hyperbolic functions into the trigonometric ones.
For the case II the $L$ operator used in \cite{AT} has the form
(\ref{eq:defL}) with $Q(u,q)$ replaced by $1/\sinh q$. The corresponding $r$
matrix has the structure (\ref{eq:defr}) with
\beq
 a=0, \qquad c_{\a\b}=-\coth q_{\a\b}, \qquad
      d_{\a\b}=-\frac{1}{2\sinh q_{\a\b}}
\eeq

In order to reproduce this result
let us start with the $L$ operator and the $r$ matrix given by
(\ref{eq:defL}), (\ref{eq:defr}) and (\ref{eq:szp-II}), (\ref{eq:Q-II})
and take the limit
$$
u\longrightarrow+\infty, \qquad Q(u,q)\longrightarrow \frac{\e^q}{\sinh q}.
$$

Removing the factors $\e^{q_{\a\b}}$ from $L^\a_\b$ by means of the
similarity transform
$$
 L\longrightarrow WLW^{-1}, \qquad W^\a_\b=\d_{\a\b}\e^{-q_\a}
$$
one obtains the $L$ operator used in \cite{AT}. The modified $r$ matrix is
calculated then as previously (\ref{eq:LLmod}) with the only difference
that one has to coordinate the limits of $u$ and $v$. The simplest choice
is $u-v=\pi i/2$. It allows to avoid the pole $(u-v)^{-1}$
and to preserve the relation (\ref{eq:defr21}).
The resulting $r$ matrix coincides with the one found in \cite{AT}.

\section{Generalized Yang-Baxter equation}

It is well known that for a purely numeric $r$ matrix a sufficient
condition for the Poisson bracket defined by (\ref{eq:LL}) to satisfy
the Jacobi identity is the classical Yang-Baxter equation \cite{FT}.
\beq
 [r^{(12)},r^{(13)}]+[r^{(12)},r^{(23)}]-[r^{(13)},r^{(32)}]=0
\label{eq:CYBE}
\eeq
(in this section we omit, for the sake of brevity,  the spectral parameters
$u$, $v$, $w$ corresponding,
respectively, to the spaces 1, 2, 3 in the tensor product
${\bf C}^N\otimes{\bf C}^N\otimes{\bf C}^N$).

The natural question arises of a proper generalization of (\ref{eq:CYBE})
for the dynamical $r$ matrices. Starting with the Jacobi identity
for the triple $L^{(1)}$, $L^{(2)}$, $L^{(3)}$
\beq
 \{\{L^{(1)},L^{(2)}\},L^{(3)}\}
+ \{\{L^{(2)},L^{(3)}\},L^{(1)}\}
+ \{\{L^{(3)},L^{(1)}\},L^{(2)}\}=0
\eeq
and using (\ref{eq:LL}) to estimate the Poisson bracket of two $L$
operators we obtain the equality \cite{BV90,Maillet}
\beq
 [R^{(123)},L^{(1)}]+[R^{(231)},L^{(2)}]+[R^{(312)},L^{(3)}]=0,
\label{eq:RL}
\eeq
where
$$
 R^{(123)}\equiv  r^{(123)}-\{r^{(13)},L^{(2)}\}+\{r^{(12)},L^{(3)}\},
$$
$r^{(123)}$ being the left-hand-side of (\ref{eq:CYBE}).

Trying to satisfy (\ref{eq:RL}) with the ansatz
\beq
 R^{(123)}=[X^{(123)},L^{(2)}]-[Y^{(123)},L^{(3)}]
\label{eq:ansatz}
\eeq
we obtain the relation between $X$ and $Y$
\beq
 Y^{(123)}=X^{(312)}
\label{eq:defY}
\eeq
or, at length,
\beq
 Y^{\a_1\a_2\a_3}_{\b_1\b_2\b_3}(u,v,w)=
 X^{\a_3\a_1\a_2}_{\b_3\b_1\b_2}(w,u,v).
\label{eq:defYind}
\eeq

We are led thus to the following generalization of the
Yang-Baxter equation for dynamical $r$ matrices
\begin{eqnarray}
 [r^{(12)},r^{13)}]+[r^{(12)},r^{(23)}]-[r^{(13)},r^{(32)}]&& \nonumber \\
-\{r^{(13)},L^{(2)}\}+\{r^{(12)},L^{(3)}\} && \nonumber \\
-[X^{(123)},L^{(2)}]+[X^{(312)},L^{(3)}]&=&0
\label{eq:genCYBE}
\end{eqnarray}
involving a new object --- the $X$ matrix.

Notice that our ansatz (\ref{eq:ansatz}) for $R^{(123)}$ is more general
than that proposed in \cite{EEKT}:
$$
  R^{(123)}=[X^{(123)},L^{(2)}-L^{(3)}],
$$
$X^{(123)}$ being assumed completely symmetric with respect to the
permutations of 123.

\begin{pr}
 For the $L$ operator (\ref{eq:defL}) and the $r$ matrix (\ref{eq:defr})
the identity (\ref{eq:genCYBE}) holds with the following matrices
$X^{(123)}$ and $X^{(312)}$
\beq
 X^{(123)}(u,v,w)=-i\sumab Q(w,q_{\a\b})\left[
   -\frac{5}{8}E^{\a\a\a}_{\a\a\b}
   +\frac{1}{8}E^{\b\b\a}_{\b\b\b}
   +\frac{1}{4}E^{\a\b\a}_{\a\b\b}
   +\frac{1}{4}E^{\b\a\a}_{\b\a\b} \right],
\label{eq:defX}
\eeq
where
$$
 E^{\a_1\a_2\a_3}_{\b_1\b_2\b_3}\equiv
 E^{\a_1}_{\b_1}\otimes E^{\a_2}_{\b_2}\otimes E^{\a_3}_{\b_3},
$$
$X^{(312)}$ being given by (\ref{eq:defY}), (\ref{eq:defYind}).
\end{pr}

{\bf Proof.} The verification of (\ref{eq:genCYBE}) is quite similar  to
that of (\ref{eq:LL}) in the proof of the Proposition 1. Again, one notices
that, since the tensor
$X^{\a_1\a_2\a_3}_{\b_1\b_2\b_3}$ has nonzero components only for combinations
of six indices $\Carre{\a_1}{\a_2}{\a_3}{\b_1}{\b_2}{\b_3}$ having no more
than {\it two} different indices,
it suffices to consider
only the combinations $\Carre{\a_1}{\a_2}{\a_3}{\b_1}{\b_2}{\b_3}$
having no more than {\it three} different indices. The obvious symmetry
of (\ref{eq:defX}) with respect to the permutations of indices allows
to restrict the list of the combinations to be verified with the total of
$1+31+90=122$ items, which is, however, too much to be done by hands.
Fortunately, a computer-assisted calculation shows
that 102 of 122 equalities result,
after substitutions (\ref{eq:defL}), (\ref{eq:defacd}) and (\ref{eq:defX}),
 in the trivial identities $0=0$.
It remains to verify only quite manageable amount of 20 identities
corresponding to the combinations
$\Carre{\a_1}{\a_2}{\a_3}{\b_1}{\b_2}{\b_3}=
\Carre121111$, $\Carre112111$, $\Carre111121$, $\Carre111112$,
$\Carre121211$, $\Carre121112$, $\Carre112211$, $\Carre112121$,
$\Carre122121$, $\Carre122112$, $\Carre121122$, $\Carre112122$,
$\Carre122221$, $\Carre122212$, $\Carre112123$, $\Carre121132$,
$\Carre123112$, $\Carre123131$, $\Carre123231$, $\Carre123312$.

These 20 identities can be divided into 3 groups. The first group
includes 6 identities corresponding to the combinations
$\Carre112123$, $\Carre121132$,
$\Carre123112$, $\Carre123131$, $\Carre123231$, $\Carre123312$
which are analogous to the identities $\carre1231$ and $\carre1223$
(see the proof of the Proposition 1) and are reduced to the standard
identity (\ref{eq:3term}).

The second group includes 12  identities corresponding to the combinations
$\Carre121111$, $\Carre112111$, $\Carre111121$, $\Carre111112$,
$\Carre121211$,  $\Carre112211$,
$\Carre122121$, $\Carre122112$, $\Carre121122$, $\Carre112122$,
$\Carre122221$, $\Carre122212$, which, like the identities
$\carre1112$, $\carre1121$, $\carre1211$, and $\carre1222$ from the
quoted proof, are reduced to the second standard identity  (\ref{eq:4term}).

The remaining 2 combinations $\Carre121112$, $\Carre112121$ lead,
respectively,  to the identities
\begin{eqnarray*}
\lefteqn{\zeta(u-v)+\zeta(v)-\zeta(u-w)-\zeta(w)=}  \\
&&\frac{Q(u-v,q_{12})Q(u-w,q_{21})-Q(v,q_{21})Q(w,q_{12})}{Q(v-w,q_{21})}, \\
\lefteqn{\zeta(u-v)+\zeta(v)-\zeta(u-w)-\zeta(w)=} \\
&&\frac{Q(u-v,q_{21})Q(u-w,q_{12})-Q(v,q_{12})Q(w,q_{21})}{Q(v-w,q_{12})}
\end{eqnarray*}
which are reduced to (\ref{eq:3term}) after applying first (\ref{eq:4term})
to the left-hand side.

\section{Discussion}

We have calculated the dynamical $r$ matrix for the elliptic Calogero-Moser
model
containing the spectral parameter. We have shown that the Poisson algebra of
$L$ and $r$ is not closed: $r$ satisfies a
generalized version of the classical Yang-Baxter equation
(\ref{eq:genCYBE}) which involves a
new object --- the $X$ matrix. In fact, there might exist more
Yang-Baxter type identities involving for instance $\{X^{(123)},L^{(4)}\}$
etc.\
which might give rise to an infinite sequence of $r$ matrices of higher
dimensions.
A remarkable feature of the matrices $r$ and $X$ for the elliptic
Calogero-Moser
model is that they depend only on coordinates $q_\a$, cf.\ \cite{FA}.
Hopefully, this property will hold also for the higher $r$ matrices.

All these questions are still waiting for an investigation. To resolve them,
probably some geometrical interpretation of the $L$ operator and $r$ matrics
might be useful. For the case of trigonometric potential and the $L$ operator
containing no spectral parameter
such an interpretation in terms of the hamiltonian reduction exists
\cite{OlP81,KKS} and has been applied recently to construct
geometrically the corresponding $r$ matrix \cite{ABT}. For the cases of
elliptic
potential and the $L$ operator with spectral parameter no geometrical
interpretation is known so far.

The dynamical $r$ matrix (\ref{eq:defr}) could provide, possibly, a mean to
construct
a separation of variables for the Calogero-Moser model in the same manner as
in the case of numerical $r$ matrices for integrable magnetic chains
\cite{Skl34}. The work in this direction is in progress.

About the quantum version of the elliptic Calogero-Moser model almost
nothing is known so far, except a construction of commutative quantum
integrals of motion \cite{OlP83}.
A possible way to approach quantization of the model could be to look for
a quantum version of the relations (\ref{eq:LL}) and (\ref{eq:genCYBE}).

A natural generalization of the Calogero-Moser model is its relativistic
version \cite{Ruijs87}. In this case, the Poisson bracket of $L$ operators
can also be represented in an $r$ matrix form, though the right-hand-side is
now quadratic in $L$. The $r$ matrices again depend only on the coordinates
$q_\a$, see \cite{BB} and \cite{Skl?} for the cases, respectively, of the $L$
operator without, and with the spectral parameter.

\bigskip
\noindent{\it Acknowledgements.}
I am grateful to J.~Avan, O.~Babelon, J.-M.~Maillet and M.~Talon for valuable
discussions.
I thank the Laboratory of Theoretical Physics of the University Paris-VI  for
hospitality.

\end{document}